\documentclass[prl,amsmath,amssymb,twocolumn,showpacs,aps,floatfix]{revtex4}
\usepackage{amsmath,amssymb,natbib,bm,graphicx,url,epsfig}
\usepackage[ansinew]{inputenc}

\newcommand{\be}{\begin{equation}}
\newcommand{\ee}{\end{equation}}
\newcommand{\bes}{\begin{equation}\begin{split}}
\newcommand{\ees}{\end{split}\end{equation}}

\begin{document}

\title{Crossover from weak localization to Shubnikov-de Haas oscillations
in a  high mobility  2D electron gas}

\author{T. A. Sedrakyan and M. E. Raikh}

\affiliation{ Department of Physics, University of Utah, Salt Lake
City, UT 84112}

\begin{abstract}
We study the magnetoresistance, $\delta\rho_{xx}(B)/\rho_0$, of a
high-mobility  2D electron gas in the domain of magnetic fields,
$B$, intermediate between the weak localization and the
Shubnikov-de Haas oscillations, where  $\delta\rho_{xx}(B)/\rho_0$
is governed by the interaction effects. Assuming short-range
impurity scattering, we demonstrate that in the {\em second order}
in the interaction parameter, $\lambda$,  a {\em linear}
$B$-dependence, $\delta\rho_{xx}(B)/\rho_0\sim
\lambda^2\omega_c/E_F$ with {\em temperature-independent} slope
emerges in this domain of $B$ (here $\omega_c$ and $E_F$ are the
cyclotron frequency and the Fermi energy, respectively). Unlike
previous mechanisms, the linear magnetoresistance is {\em
unrelated} to the electron executing the full Larmour circle, but
rather originates from the impurity scattering via the
$B$-dependence of the {\em phase} of the impurity-induced Friedel
oscillations.

\end{abstract}


\maketitle

{\noindent \it Introduction.} There are two prominent regimes of
low-temperature magnetotransport in a 2D electron gas: weak
localization \cite{hikami80} and Shubnikov-de Haas oscillations.
Weak localization correction dominates magnetoconductivity at low
fields, $\omega_c\tau < \omega_c^{tr}\tau$,
where $\tau$ is the impurity scattering time. Characteristic
frequency, $\omega_c^{tr}$, is determined from the condition
\cite{dyakonov94} that the magnetic flux through a triangle with a
side of a mean free path, $l=v_{\mbox{\tiny F}}\tau$, is equal to
the flux quantum, which yields $\omega_c^{tr}\tau =(k_{\mbox{\tiny
F}}l)^{-1}$.
Here $v_{\mbox{\tiny F}}$ and $k_{\mbox{\tiny F}}$ are the Fermi
velocity and Fermi momentum, respectively. On the other hand, the
oscillatory in
$B$ corrections to the resistivity,
$\delta\rho_{xx}(B)=\rho_{xx}(B)-\rho_0$,  where
$\rho_0=\sigma_0^{-1}=\rho_{xx}(0)=h/e^2(k_{\mbox{\tiny F}}l)$,
develop at high fields,  $\omega_c\tau \gtrsim 1$. Thus, the
boundaries between the low-field and the high-field regimes are
separated by a large parameter, $k_{\mbox{\tiny F}}l$.

The behavior of $\delta \rho_{xx}(B)$ in the crossover regime,
$\omega_c^{tr} \tau < \omega_c\tau <1$ has been studied
experimentally for more than two decades, see, {\em e.g.}, Refs.
\cite{tsui,group}.
It is commonly
accepted that this behavior is governed by the interaction
effects. More specifically, the $B$-dependence of $\delta
\rho_{xx}$ is believed to come exclusively from the inversion of
the conductivity tensor \cite{houghton82}
\begin{eqnarray}
\label{inversion} \delta\rho_{xx}^{int}(B,T)\approx
\rho_0^2\left(\omega_c^2\tau^2-1\right)\delta \sigma_{xx}^{int}(T)
\end{eqnarray}
where $\delta \sigma_{xx}^{int}(T)$ is the {\em zero-field}
interaction correction \cite{AAL} to the conductance. This
correction is derived under assumption that, in course of an
electron-electron collision, the  electron
performs many steps $\sim l$
of diffusion;
for $\omega_c\tau < 1$ the
orbital  effect of $B$
on each step  is neglected.

In experiments \cite{tsui,group}
the electron
mobilities were relatively low, so that $k_{\mbox{\tiny F}}l$ was
$\lesssim 10$. In the present paper we demonstrate that  for very
big values of $k_{\mbox{\tiny F}}l\gg 1$, like in Refs.
\cite{zudov01',zudov01,mani02}, the {\em higher-order}
electron-electron interaction processes, at distances $\lesssim l$
are strongly sensitive to  $B$ even for $\omega_c\tau <1$. Due to
these processes, each involving {\em two scattering acts}, that
were neglected in previous considerations, a lively $B$-dependence
of $\delta\sigma_{xx}$ emerges in the crossover domain
$\omega_c^{tr} < \omega_c <\tau^{-1}$. This dependence, in turn,
translates into the  $B$-dependence of $\delta\rho_{xx}$, which is
much stronger than the one coming from the inversion of the
conductivity tensor. Namely, we find the interaction contribution
to $\sigma_{xx}$ in the form
\begin{equation}
\label{contribution}
\frac{\delta\sigma_{xx}(B)}{\sigma_0}=\frac{4\lambda^2}{(k_{\mbox{\tiny
F}}l)^{3/2}} \;\text{\Large
F}_1\left(\frac{\omega_c}{\Omega_l}\right),~~~\Omega_l\tau=(k_{\mbox{\tiny
F}}l)^{-1/2},
\end{equation}
where $\lambda$ is the dimensionless interaction constant.
It is important that the characteristic field, $\Omega_l$, {\em
lies in the crossover domain}, i.e.,  it is much bigger than
$\omega_c^{tr}$, but much smaller than $1/\tau$.

The function $\text{\large F}_1$ (Fig.~3) has the following
asymptotes
\begin{eqnarray}
\label{lowT} \text{\Large F}_1(x)=\left\{
\begin{array}{cc}
-x^2/8 ,& x\ll 1\qquad\text{(a)}\\
-2x/3,& x\gg 1.\qquad\text{(b)}
\end{array}\right.
\end{eqnarray}
The new scale of the cyclotron frequencies, $\Omega_l$, originates
from the new physical process: {\em double backscattering} from
the impurity-induced Friedel oscillations, see Figs. 1 and 2. By
virtue of the fact that this process causes the $B$-dependence of
the electron scattering time, the correction
Eq.~(\ref{contribution}) enters also into magnetoresistance,
$\delta\rho_{xx}(B)/\rho_0$. This magnetoresistance is much
stronger than $\omega_c^2\tau^2\delta\sigma_{xx}^{int}(T)$,
defined by Eq.~(\ref{inversion}). Indeed, within a logarithmic
factor, $\delta\sigma^{int}/\sigma_0 \sim \lambda\;(k_{\mbox{\tiny
F}}l)^{-1}$. Then it follows from
Eqs.~(\ref{inversion})-(\ref{lowT}) that
\begin{eqnarray}
\label{ratio} \frac{\delta\rho _{xx}}{\delta\rho _{xx}^{int}}\sim
\left\{
\begin{array}{cc}
\lambda\; (k_{\mbox{\tiny F}}l)^{1/2},& \;\; (k_{\mbox{\tiny F}}l)^{-1}<\omega_c\tau<(k_{\mbox{\tiny F}}l)^{-1/2}\\
\lambda\;(\omega_c\tau)^{-1},& \;\;(k_{\mbox{\tiny
F}}l)^{-1/2}<\omega_c\tau<1.
\end{array}\right.
\end{eqnarray}
We see that in both limits the ratio Eq.~(\ref{ratio}) is big.

Up to now we considered only low-$T$ behavior of
magnetoresistance. With increasing mobility, the condition $T\tau
>1$ is met even at low temperatures. Under this condition,
the ballistic correction \cite{dolgopolov,Narozhny}
$\delta\sigma_{xx}^{int}(T)/\sigma_0 \sim \lambda T/E_{\mbox{\tiny
F}}$ is the leading temperature correction to $\delta\sigma_{xx}$.
Its origin is the interference between the impurity  scattering
and the scattering from the Friedel oscillation; linear
$T$-dependence results from the fact that, in the ballistic
regime, the spatial extent of the Friedel oscillations is limited
by the length  $r_{\mbox{\tiny T}}=v_{\mbox{\tiny F}}/2\pi T$
rather than by $l$. Since the ballistic correction is merely a
$B$-independent renormalization of $\tau$, it does not contribute
to $\delta \rho_{xx}$.
Instead \cite{Mirlin1},
the dependence $\delta\rho_{xx}^{int}(B)$
comes from a small $B$-dependent portion, $\sim \omega_c^2/T^2$,
of $\delta\sigma_{xx}^{int}(T)$ yielding
$\delta\rho_{xx}^{int}/\rho_0 \sim
\lambda\omega_c^2/E_{\mbox{\tiny F}}T$.
\begin{figure}[t]
\centerline{\includegraphics[width=85mm,angle=0,clip]{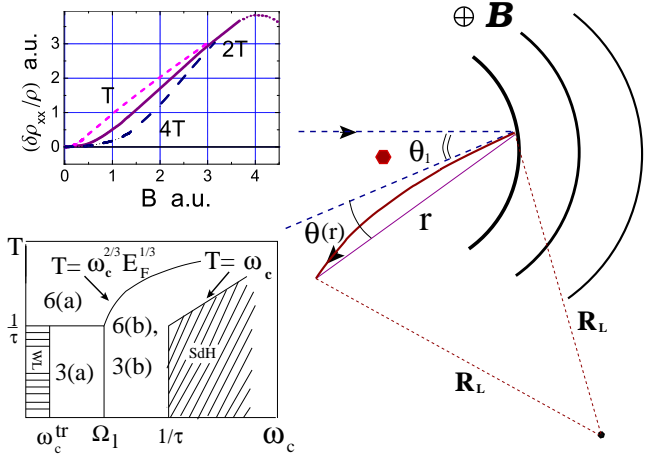}}
\caption{Schematic illustration of electron backscattering from
the Friedel oscillation (arcs), created by the short-range
impurity (big dot). Magnetic field causes an additional deflection
by the angle $\theta_{\mbox{\tiny B}}(r)\approx r/R_{\mbox{\tiny
L}}$ due to the trajectory curving and the resulting additional
phase $\Psi_{\mbox{\tiny B}}(r)=k_{\mbox{\tiny
F}}r^3/24R_{\mbox{\tiny L}}^2$. Lower inset: domains of different
behaviors of $\rho_{xx}$ on the $B$-$T$ plain are shown
schematically. Upper inset: evolution of ballistic
magnetoresistance with increasing temperature;
$\delta\rho_{xx}(B)$ dependencies are plotted from
Eqs.~(\ref{contribution2}) and (\ref{Toriginal}) for three
temperatures: $T$, $2T$, and $4T$; Dotted line illustrates a
crossover, Eq.~(\ref{general}), from positive to negative
magnetoresitance. }
\end{figure}

Due to the cutoff at distances $\sim r_{\mbox{\tiny T}}$,  our
result Eq.~(\ref{contribution}) in the ballistic regime assumes
the form
\begin{equation}
\label{contribution2}
\frac{\delta\sigma_{xx}(B)}{\sigma_0}\!=\!4\lambda^2
\left(\frac{\pi T}{E_{\mbox{\tiny F}}}\right)^{3/2} \text{\Large
F}_2\left(\frac{\omega_c}{2\pi^{3/2}\Omega_{\mbox{\tiny
T}}}\right),~~\Omega_{\mbox{\tiny
T}}=\frac{T^{3/2}}{E_{\mbox{\tiny F}}^{1/2}},
\end{equation}
with characteristic ``ballistic'' cyclotron frequency,
$\Omega_{\mbox{\tiny T}}$,
much smaller than the temperature. The asymptotes of the
dimensionless function $\text{\large F}_2$
are the following
\begin{eqnarray}
\label{highT} \text{\Large F}_2(x)=\left\{
\begin{array}{cc}
-0.7x^2,& x\ll 1\qquad \text{(a)}\\
-2x/3,& x\gg 1.\qquad \text{(b)}
\end{array}\right.
\end{eqnarray}
Comparison of the corresponding correction to $\rho_{xx}$ with
$\delta\rho_{xx}^{int}$ from Ref. \cite{Mirlin1} yields
\begin{eqnarray}
\label{ratio1} \Biggl(\frac{\delta\rho _{xx}}{\delta\rho
_{xx}^{int}}\Biggr)_{T\tau>1}\sim \left\{
\begin{array}{cc}
\lambda\;\bigl(E_{\mbox{\tiny F}}/T\bigr)^{1/2},&
\;\;\omega_c<\Omega_{\mbox{\tiny
T}}<T \\
\lambda\;\bigl(T/\omega_c\bigr),& \;\;\Omega_{\mbox{\tiny
T}}<\omega_c<T.
\end{array}\right.
\end{eqnarray}
For $\lambda \sim 1$ both ratios are big either in parameter
$E_{\mbox{\tiny F}}/T$ or in
$T/\omega_c$, the latter
ensures that Shubnikov-de Haas oscillations are smeared out even
in the ballistic regime.

The fact  that the interaction correction
Eq.~(\ref{contribution2}) comes from short distances, $\sim
r_{\mbox{\tiny T}}$ suggests that $\omega_c\tau$ may be both,
smaller or larger than $1$, in the ballistic regime, see Fig.~1,
inset. Therefore, one has to use Eq.~(\ref{inversion}) to
transform $\delta\sigma_{xx}(B)$ into magnetoresistance. Then in
the ``strong-field'' domain, $\Omega_{\mbox{\tiny T}}<\omega_c<T$,
we find from  Eq.~(\ref{contribution2})
\begin{eqnarray}
\label{general}
\delta\rho_{xx}/\rho_0=(4/3)\;\lambda^2\left(1-\omega_c^2\tau^2\right)\left(\omega_c/E_{\mbox{\tiny
F}}\right),
\end{eqnarray}
i.e., positive magnetoresistance crosses over to negative at
$\omega_c\tau=3^{-1/2}$. Below we demonstrate the emergence of the
new $\omega_c$-scales, $\Omega_l$ and $\Omega_{\mbox{\tiny T}}$,
qualitatively.
\begin{figure}[t]
\centerline{\includegraphics[width=90mm,angle=0,clip]{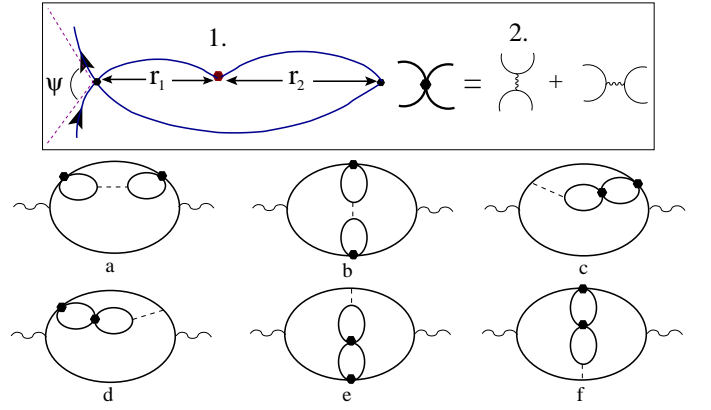}}
\caption{Diagrams for the second-order (in the interaction
strength, $\lambda$) correction, $\delta\sigma_{xx}(B)$, to the
magnetoconductivity. Diagram $a$ describes combined {\em double}
scattering from the impurity (big dot) and from the Friedel
oscillation; this process is also illustrated in the inset 1,
where $\alpha\approx[\theta_1+\theta_{\mbox{\tiny
B}}(r_{\mbox{\tiny T}})]$ is the net scattering angle, see the
text. Two types of four-leg interaction vertices are combined into
dots (inset 2).}
\end{figure}

{\noindent \it Qualitative derivation of
Eqs.~(\ref{contribution}), (\ref{contribution2}).} Consider first
high temperatures, $T\tau > 1$. We will follow the efficient line
of reasoning of Refs. \cite{Narozhny,Mirlin1,Mirlin2}, which is
based on the analysis of the expression for transport scattering
time
\begin{eqnarray}
\label{rate}
 \tau^{-1}=\int d\Theta/2\pi(1-\cos\Theta)\vert f(\Theta)\vert^2,
\end{eqnarray}
where $f(\Theta)$ is the {\em full} scattering amplitude,
$f_0(\Theta)+f_1(\Theta)$, from the impurity and the
impurity-induced potential. Assume a short-range impurity
potential, $U_{imp}(r)$. In the first order in interaction
strength and for scattering angle $\pi-\Theta =\theta_1 \ll 1$
(see Fig.~1.) the amplitude $f_1$ is given by
\begin{eqnarray}
\label{scatt1} f_1(\theta_1,T)=-\lambda
g\int_0^{\infty}\frac{dr}{r}\sin\left(2k_{\mbox{\tiny F}}r\right)
A\left(\frac{r}{r_{\mbox{\tiny T}}}\right)\nonumber\\ \times
J_0\Biggl(2k_{\mbox{\tiny F}}r\left[1-\frac{\theta
_1^2}{2}\right]\Biggr),
\end{eqnarray}
where $J_0$ is the Bessel function of zero order, $g=\int d{\bf
r}U_{imp}(r)$, and the function $A(x)=x/\sinh x$ is the spatial
temperature damping factor (see {\em e.g.,} \cite{Narozhny}). It
follows from Eq.~(\ref{scatt1}) that the characteristic angular
interval for the enhanced backscattering is $\theta_1\sim
(k_{\mbox{\tiny F}}r_{\mbox{\tiny T}})^{-1/2}$. On the other hand,
the relative magnitude of enhancement can be estimated from
Eq.~(\ref{scatt1}) as $\left[f_1(0,T)-f_1(0,0)\right]\sim \lambda
f_0(k_{\mbox{\tiny F}}r_{\mbox{\tiny T}})^{-1/2}$. Thus, the
relative $T$-dependent correction to $\tau^{-1}$ and,
correspondingly, to $\sigma_{xx}$, is $\sim
\left(\lambda/k_{\mbox{\tiny F}}r_{\mbox{\tiny T}}\right)\sim
\lambda T/E_{\mbox{\tiny F}}$, as in
Refs.~\cite{dolgopolov,Narozhny}.


According to Ref. \cite{Mirlin1}, incorporating magnetic field
into the above picture amounts to adding to the scattering angle,
$\theta_1$, the  angle, $\theta_{\mbox{\tiny B}} (r_{\mbox{\tiny
T}}) \sim r_{\mbox{\tiny T}}/R_{\mbox{\tiny L}}$, which accounts
for the fact that, upon travelling a distance, $r$, in magnetic
field, the electron experiences angular deflection by
$\theta_{\mbox{\tiny B}}(r)=r/R_{\mbox{\tiny L}}$, see Fig.~1.
Here $R_{\mbox{\tiny L}}=v_{\mbox{\tiny F}}/\omega_c$ is the
Larmour radius.
In Ref.~\cite{Mirlin1} the modification of the amplitude, $f_1$,
by magnetic field is neglected. Then the effect of $B$ on the
scattering rate Eq.~(\ref{rate}) reduces to the correction $\sim
-\lambda\left[\theta_{\mbox{\tiny B}}(r_{\mbox{\tiny
T}})\right]^2/\tau$; the factor $\left[\theta_{\mbox{\tiny
B}}(r_{\mbox{\tiny T}})\right]^2$ comes from integrating
$\left[1+\cos(\theta_1+\theta_{\mbox{\tiny B}}(r_{\mbox{\tiny
T}}))\right]$ over $\theta_1$. By noting that
$\left[\theta_{\mbox{\tiny B}}(r_{\mbox{\tiny T}})\right]^2\sim
\omega_c^2/T^2$, we reproduce the result of Ref.~\cite{Mirlin1}
for $\delta\rho_{xx}^{int}(B)$.

The new scale, $\Omega_{\mbox{\tiny T}}$, introduced in Eq.~
(\ref{contribution2}), can be now inferred from the condition,
$\theta_{\mbox{\tiny B}}(r_{\mbox{\tiny T}}) < \theta_1$, that the
replacement $\theta_1\rightarrow (\theta_1+\theta_{\mbox{\tiny
B}}(r_1))$ in the integrand of Eq.~(\ref{scatt1})
does not change the
amplitude, $f_1$. Indeed, equating $\theta_{\mbox{\tiny
B}}(r_{\mbox{\tiny T}})$ to $\theta_1\sim (k_{\mbox{\tiny
F}}r_{\mbox{\tiny T}})^{-1/2}$, we find
$\omega_c\!=\!T^{3/2}/E_{\mbox{\tiny F}}^{1/2}\!\sim\!
\Omega_{\mbox{\tiny T}}$.

It might seem that in the opposite case, $\theta_{\mbox{\tiny
B}}(r_{\mbox{\tiny T}})
>\theta_1$, the size of the scattering region would be determined
by the magnetic phase, $\Psi_{\mbox{\tiny B}}(r)$, see Fig. 1
caption, as
\begin{equation}
\!\Psi_{\mbox{\tiny B}}(r_{\mbox{\tiny B}})=\left(k_{\mbox{\tiny
F}}r_{\mbox{\tiny B}}^3/24R_{\mbox{\tiny L}}^2\right)\sim
1,~{\text{{ i.e.,}}}~r_{\mbox{\tiny B}}\sim (R_{\mbox{\tiny
L}}^2/k_{\mbox{\tiny F}})^{1/3},
\end{equation}
rather than by $r_{\mbox{\tiny T}}$. This, however, is not the
case. The reason is that the rigorous treatment \cite{we} requires
incorporating the magnetic phase, $-2\Psi_{\mbox{\tiny B}}(r)$,
not only into the argument of the Bessel function in
Eq.~(\ref{scatt1}) but into  the argument of sine as well. The
latter describes field-induced modification of the Friedel
oscillations \cite{we}.
As a result, the $B$-dependent phase factors {\em cancel out}.

Our main point is that the cancellation {\em does not occur} in
the second-order process in the interaction strength. As
illustrated in Fig.~2 (inset 1),  the  backscattering is the
result of {\em two} virtual scattering processes from the Friedel
oscillation. The contribution to the scattering amplitude from
this process reads (see also inset 1 in Fig. 2)
\begin{eqnarray}
\label{second} \tilde{f}_1(\theta_1)=\frac{\lambda ^2
g}{2\pi}\int\frac{d{r}_1d{r}_2d\varphi_{r_1}}{r_1r_2}
A\left(\frac{r_1}{r_{\mbox{\tiny T}}}\right)A\left(\frac{r_2}{r_{\mbox{\tiny T}}}\right)\qquad\\
\times\sin\bigl(2k_{\mbox{\tiny F}}r_1\bigr)
J_0\Biggl(2k_{\mbox{\tiny F}}\vert{\bf r}_1-{\bf
r}_2\vert\left[1-\frac{\theta_1^2}{2}\right]\Biggr)
\sin\bigl(2k_{\mbox{\tiny F}}r_2\bigr). \nonumber
\end{eqnarray}
It is seen from Eq.~(\ref{second}) that the characteristic value
of the angle, $\pi-\varphi_{r_1}$, between ${\bf r}_1$ and ${\bf
r}_2$ is $\sim (k_{\mbox{\tiny F}}r_1)^{-1/2}$. With magnetic
phase
$\Psi_{\mbox{\tiny B}}(r)=\left(k_{\mbox{\tiny
F}}r^3/24R_{\mbox{\tiny L}}^2\right)$ included in the arguments of
sines and the Bessel function, the slow oscillating term in the
integrand of Eq.~(\ref{second}) will acquire the form
$-\sin\left[\Phi_{\mbox{\tiny B}}(r_1,r_2)-k_{\mbox{\tiny F}}(r_1
+ r_2)\theta_1^2+\pi/4\right]$, where
\begin{eqnarray}
\Phi_{\mbox{\tiny B}}(r_1,r_2)\!\!&=&\!\!
2\Psi_{\mbox{\tiny B}}(r_1)+2\Psi_{\mbox{\tiny B}}(r_2)
-2\Psi_{\mbox{\tiny B}}(r_1+ r_2)~~~~
\\
&=&\!\! - k_{\mbox{\tiny F}}r_1r_2(r_1 + r_2)/4R_{\mbox{\tiny
L}}^2.\nonumber
\end{eqnarray}
We are now in position to estimate the $\lambda^2$-correction to
the scattering rate Eq.~(\ref{rate}) in both domains $\omega_c
<\Omega_{\mbox{\tiny T}}$ and $\omega_c > \Omega_{\mbox{\tiny
T}}$. For low magnetic field, both $\varphi_{r_1}$ and $\theta_1$
are $\sim (k_{\mbox{\tiny F}}r_{\mbox{\tiny T}})^{-1/2}$. The
integral in Eq.~(\ref{second}) can be estimated as
$[\tilde{f}_1(\theta_1,B)-\tilde{f}_1(\theta_1,0)] \sim
\lambda^2\varphi_{r_{\mbox{\tiny T}}}(k_{\mbox{\tiny
F}}r_{\mbox{\tiny T}})^{-1/2}\Phi_{\mbox{\tiny B}}(r_{\mbox{\tiny
T}},r_{\mbox{\tiny T}})$. Then the integration over $\theta_1$ in
Eq.~(\ref{rate}) would yield the relative $B$-dependent correction
$\sim (k_{\mbox{\tiny F}}r_{\mbox{\tiny
T}})^{-3/2}\Phi_{\mbox{\tiny B}}(r_{\mbox{\tiny T}},r_{\mbox{\tiny
T}})\sim \lambda^2\omega_c^2/(T^{3/2}E_{\mbox{\tiny F}}^{1/2})$ to
the scattering rate. This leads to the estimate
$\delta\rho_{xx}(B)/\rho_0 \sim
\lambda^2\omega_c^2/(T^{3/2}E_{\mbox{\tiny F}}^{1/2})$, which
coincides with our Eq.~(\ref{highT}). For high magnetic fields we
have $\varphi_{r_1}\sim \theta_1\sim (k_{\mbox{\tiny
F}}r_{\mbox{\tiny B}})^{-1/2}$; the difference
$[\tilde{f}_1(\theta_1,B)-\tilde{f}_1(\theta_1,0)]$ is now $\sim
(k_{\mbox{\tiny F}}r_{\mbox{\tiny B}})^{-1}\Phi_{\mbox{\tiny
B}}(r_{\mbox{\tiny B}},r_{\mbox{\tiny B}})$, so that the estimate
for $\delta\rho_{xx}(B)/\rho_0$ assumes the form
$\lambda^2\omega_c/E_{\mbox{\tiny F}}$ again in accord with
Eq.~(\ref{highT}). Note, that ``strong-field''magnetoresistance in
the domain $\Omega_{\mbox{\tiny T}}<\omega_c<T$ is {\em
temperature-independent} (see upper inset in Fig.~1).

Consideration for low temperatures leading to Eq.~(\ref{lowT}) is
absolutely similar. On the quantitative level, one has to replace
the temperature damping factor $A(r/r_{\mbox{\tiny T}})$ by the
probability $\exp(-2r/l)$ that electron does not encounter other
impurity in course of scattering from a given impurity and from
the Friedel oscillations, created by it.

{\noindent \it Outline of the derivation.} It is most convenient
to calculate the magnetoconductivity, $\delta\sigma_{xx}(B)$, in
the coordinate space. In the {\bf r}-space,  Friedel oscillation
manifests itself via a  polarization operator, $\Pi(r,B)$ , which
has the following form \cite{we}
\begin{eqnarray}
\label{operator} \Pi_{\omega}({\bf
r},0)=-\frac{\pi\nu_0^2\hbar^4}{2k_{\mbox{\tiny
F}}r}\Biggl[i\vert\omega\vert+ \frac{v_{\mbox{\tiny
F}}}{r}\sin\left(2k_{\mbox{\tiny
F}}r-\frac{\omega_c^2E_{\mbox{\tiny F}}\;r^3}{6v_{\mbox{\tiny
F}}^3}\right)
\nonumber\\
\times A\Biggl(\frac{r}{r_{\mbox{\tiny
T}}}\Biggr)\Biggr]\exp\Biggl\{\frac{i\vert\omega\vert
r}{v_{\mbox{\tiny F}}}-\frac{r}{l}\Biggr\},\qquad
\end{eqnarray}
where, $\nu_0$ is the 2D density of states. The $B$-dependent term
in the argument of sine coincides within a numerical factor with
magnetic phase, $k_{\mbox{\tiny F}}r\left[\theta_{\mbox{\tiny
B}}(r)\right]^2$, derived above. Diagram $a$ in Fig.~2 contains
two polarization bubbles connected by an impurity line, and
positioned in such a way, that they play the role of an effective
scatterer. Then the entire diagram $a$ describes the contribution
to $\sigma_{xx}$ from the {\em double} scattering from the Friedel
oscillations. Analytical expression for this diagram in terms of
the polarization operator Eq.~(\ref{operator}) is the following
\begin{eqnarray}
\label{expression} \frac{\delta\sigma_{xx}(B)}{\sigma_0}=
\frac{\lambda^2}{\pi\nu_0^4} \int\!\! d{\bf r}_1\!\int\!\! d{\bf
r}_2\;\left[\frac{1}{\omega}\;\text{Im}\Pi_{\omega}({\bf r}_1,{\bf
r}_2)\right]_{\omega\rightarrow
0}\nonumber\\
\times\text{Re}\;\Bigl\{\Pi_0(0, {\bf r}_1)\Pi_0({\bf
r}_2,0)\Bigr\},\qquad
\end{eqnarray}
where we assumed that the interaction is short-ranged,
$V(q)\approx \text{const}(q)=V_0$ \cite{footnote}, so that
$\lambda=\nu_0V_0$. Our ``low-temperature'' result
Eq.~(\ref{contribution}) emerges upon substitution
Eq.~(\ref{operator}) into Eq.~(\ref{expression}), setting
$A(r/r_{\mbox{\tiny T}})=1$, extracting a slow term from three
rapidly oscillating sines
and, finally, performing integration over the azimuthal positions,
$\varphi_{{\bf r}_1}$, $\varphi_{{\bf r}_2}$ of the points ${\bf
r}_1$, ${\bf r}_2$. To arrive to our ballistic result
Eq.~(\ref{contribution2}), one should keep $A(r/r_{\mbox{\tiny
T}})$ in Eq.~(\ref{operator}) and take the limit $l\rightarrow
\infty$. The final form of the dimensionless functions
$\text{\large F}_1(x)$, $\text{\large F}_2(x)$ is the following
\begin{eqnarray}
\label{conduct}&&\!\!\!\!\!\!\!\text{\large
F}_1(x)\!=\!\frac{1}{\pi^{3/2}}\!\int\limits_{\rho_1>\rho_2}\!\!\!\frac{d\rho_1d\rho_2}{(\rho_1\rho_2)^{3/2}}
\bigl\{
\mathcal{H}^{-}(\rho_1,\rho_2,x)e^{-2\rho_1}\nonumber\\&&\qquad\qquad\qquad\qquad+
\mathcal{H}^{+}(\rho_1,\rho_2,x)e^{-2(\rho_1+\rho_2)}\bigr\},\\
\label{Toriginal} &&\!\!\!\!\!\!\!\text{\large
F}_2(x)\!=\!\frac{1}{\pi^{3/2}}\!\int\limits_{\rho_1>\rho_2}\!\!\!\frac{d\rho_1d\rho_2}{(\rho_1\rho_2)^{3/2}}\bigl\{
\mathcal{H}^{-}(\rho_1,\rho_2,x)A\left(\rho_1\right)
A\left(\rho_2\right)\times\nonumber\\&&\!\!\!\!\!\!\!
A\left(\rho_1\!-\!\rho_2\right)+
\mathcal{H}^{+}(\rho_1,\rho_2,x)A\left(\rho_1\right)
A\left(\rho_2\right) A\left(\rho_1\!+\!\rho_2\right)
 \bigr\},
\end{eqnarray}
where $\mathcal{H}^{\pm}(\rho_1,\rho_2,x)=
(\rho_1\pm\rho_2)^{-1/2}\{\sin(\pi/4)-\sin[x^2\;\rho_1\rho_2(\rho_1\pm\rho_2)+\pi/4]\}$,
and $\pm$ corresponds to ${\bf r}_1$, ${\bf r}_2$ on the same and
opposite sides from ${\bf r}=0$, respectively.

Qualitative derivation pertained to the diagram $a$ in Fig.~2.
There are  however other virtual, second-order in
$\lambda$, processes that give rise to the contributions to
$\delta\sigma_{xx}(B,T)$, similar to Eq.~(\ref{expression}). For
example, the relevant $\lambda^2$ term can come not only from the
double
 backscattering
of an electron by  Friedel oscillation with magnitude $\lambda$
but also from a direct scattering from an impurity and from
``convolution'' of the two Friedel oscillations (diagram $c$ in
Fig.~2) $\propto\lambda^2\int d{\bf r_1}\bigl[\sin
\left(2k_{\mbox{\tiny F}}\vert {\bf r}-{\bf r}_1
\vert\right)/\vert {\bf r}-{\bf r}_1\vert^2\bigr]\; \left[\sin
\left(2k_{\mbox{\tiny F}}r_1\right)/r_1^2\right]$. Important is
that all contributions $\sim \lambda^2$ differ only by a numerical
factor. Resulting combinatorial factor, $32$, is reflected in
Eqs.~(\ref{contribution}), and (\ref{contribution2}).

In Fig.~3  we show functions $\text{{\large F}}_1(x)$ and
$\text{{\large F}}_2(x)$ calculated numerically from
Eqs.~(\ref{conduct}), (\ref{Toriginal}).
Magnetoresistance is related to $\text{{\large F}}_{1,2}$ via
additional factor $(\omega_c^2\tau^2 -1)$.
In  accord with  qualitative analysis, both functions
are quadratic for $x\ll 1$ and linear
for $x\gg 1$.
\begin{figure}[t]
\centerline{\includegraphics[width=65mm,angle=0,clip]{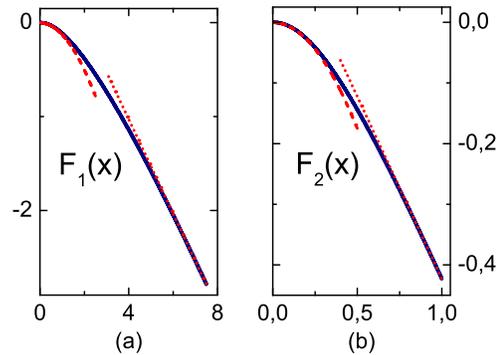}}
\caption{a)
Magnetoconductivity,
$({\delta\sigma_{xx}}/{\sigma_0})[{(k_{\mbox{\tiny
F}}l)^{3/2}}/{4\lambda^2}]$, at low $T$ is plotted from
Eq.~(\ref{conduct}) vs. dimensionless magnetic field
$x=\omega_c/\Omega_l$; b) Ballistic magnetoconductivity
$({\delta\sigma_{xx}}/{\sigma_0})[{E_{\mbox{\tiny
F}}^{3/2}}/{4\lambda^2T^{3/2}}]$ is plotted from
Eq.~(\ref{Toriginal}) vs. dimensionless magnetic field
$x=\omega_c/(2\pi^{3/2}\Omega_{\mbox{\tiny T}})$. Dashed lines for
low fields are the $x\ll 1$ asymptotes in Eqs.~(\ref{lowT}),
(\ref{highT}). Dotted lines illustrate  linear behavior of
$\delta\sigma_{xx}$ at $x\gtrsim 1$.}
\end{figure}

{\noindent \it  Discussion and estimates.} Our main result is a
novel scale  of magnetic fields, $\omega_c\tau=(k_{\mbox{\tiny
F}}l)^{-1/2}$, and a linear magnetoresistance
$\delta\rho_{xx}(B)/\rho_0\sim \lambda^2\omega_c/E_{\mbox{\tiny
F}}$
within the interval $(k_{\mbox{\tiny F}}l)^{-1/2}<\omega_c\tau<1$.
In the samples with moderate mobility \cite{tsui,group}
$\mu\sim 10^4$cm$^2$/V~s this interval is
narrow, $(k_{\mbox{\tiny F}}l)^{-1/2}\approx 0.3$ for $n=2\cdot
10^{11}$cm$^{-2}$ and
$\delta\rho_{xx}(B)$-dependencies in \cite{tsui,group}
are indeed
weak and quadratic in the crossover region. By contrast, the data
in Refs.~\cite{zudov01',zudov01,mani02} for
$\mu\gtrsim 2\cdot 10^6$cm$^2$/V~s exhibit extended intervals of
$B$, from $0.02$ Tesla to $0.2$ Tesla, in which $\delta\rho_{xx}$
is strong and linear with either positive or negative slopes. Our
theory predicts linear $\delta\rho_{xx}(B)$ only for
$\omega_c\tau<1$, which was not the case in the above domain of
$B$.
Throughout the paper we assumed that disorder is short-range. For
smooth disorder there exists a specific  regime of ballistic
magnetotransport, $T\tau>1$, where  Shubnikov-de Haas oscillations
are suppressed, i.e.,  $T>\omega_c$, but the field is strong,
$\omega_c\tau>1$. As was demonstrated in Ref. \cite{Mirlin2} and
confirmed experimentally in Ref. \cite{Savchenko},
magnetoresistance,
$\delta\rho_{xx}/\rho_0\sim \lambda(\omega_c\tau)^2(k_{\mbox{\tiny
F}}l)^{-1}(T\tau)^{-1/2}$ in this regime has a distinct
$T$-dependence. However, the $B$-dependence still comes from the
inversion of the conductivity tensor.

We gratefully acknowledge the discussions with M.~A.~Zudov and R.
R. Du.


\begin{references}
\bibitem{footnote}
Effective interaction constant is expressed through $V(q)$ as
$\lambda=\nu_0\bigl[V(0)-2V(2k_{\mbox{\tiny F}})\bigr]$, see Fig.
2, inset 2.
\end{references}
\end{document}